\documentclass[12pt,letterpaper]{article}

\usepackage{amsmath}
\usepackage{amssymb}
\usepackage{graphicx}
\usepackage{float}     
\usepackage[titletoc]{appendix}
\usepackage[hidelinks]{hyperref}  
\usepackage{authblk}  
\usepackage{lineno}   
\usepackage[export]{adjustbox}
\usepackage{caption}
\usepackage{subcaption}

\begin{document}

\pagestyle{plain}

\title{
Position Resolution of a Scintillator Hodoscope Employing Triangular Counters with
Embedded Wavelength Shifting Fibers
}
\author[1]{E. Craig Dukes}
\author[1]{Ralf Ehrlich}
\author[1]{Daniel Lee}
\author[1]{Luke Watson}
\author[2]{Alan Bross}
\author[2]{Sten Hansen}
\author[2]{Paul Rubinov}
\affil[1]{Physics Department, University of Virginia, Charlottesville, VA, 22904, USA}
\affil[2]{Fermi National Accelerator Laboratory, Batavia, IL, 60510, USA}

\maketitle

\begin{abstract}
Scintillator counters employing embedded wavelength-shifting fibers have been
used in particle physics experiments for several decades. Such counters have 
been produced with square, rectangular, and triangular profiles.
An advantage of arrays of triangular counters is that their position resolution can
be greatly enhanced by interpolation between adjacent counters using their relative
light yields. We report here on a test-beam study of the position resolution of such
a scintillator hodoscope and compare the results to a simulation. We find
an order of magnitude improvement in the position resolution over that found by
simply using the fiber separation.
\end{abstract}


\section{Introduction}
\label{intro}
We have fabricated and tested in a 120\,GeV proton beam two small scintillator hodoscopes 
employing embedded wavelength-shifting (WLS) fibers. The triangular profile of the hodoscope
counters allows the relative light yield of adjacent counters to be used to improve the position 
resolution from that determined by simply using the fiber pitch.
We report below on a simulation of the position resolution of such
a scintillator hodoscope and compare the results to a test-beam study.  

\section{The Hodoscope Design}
\label{sec:design}

\begin{figure}
\centering
\includegraphics[width=0.95\textwidth]{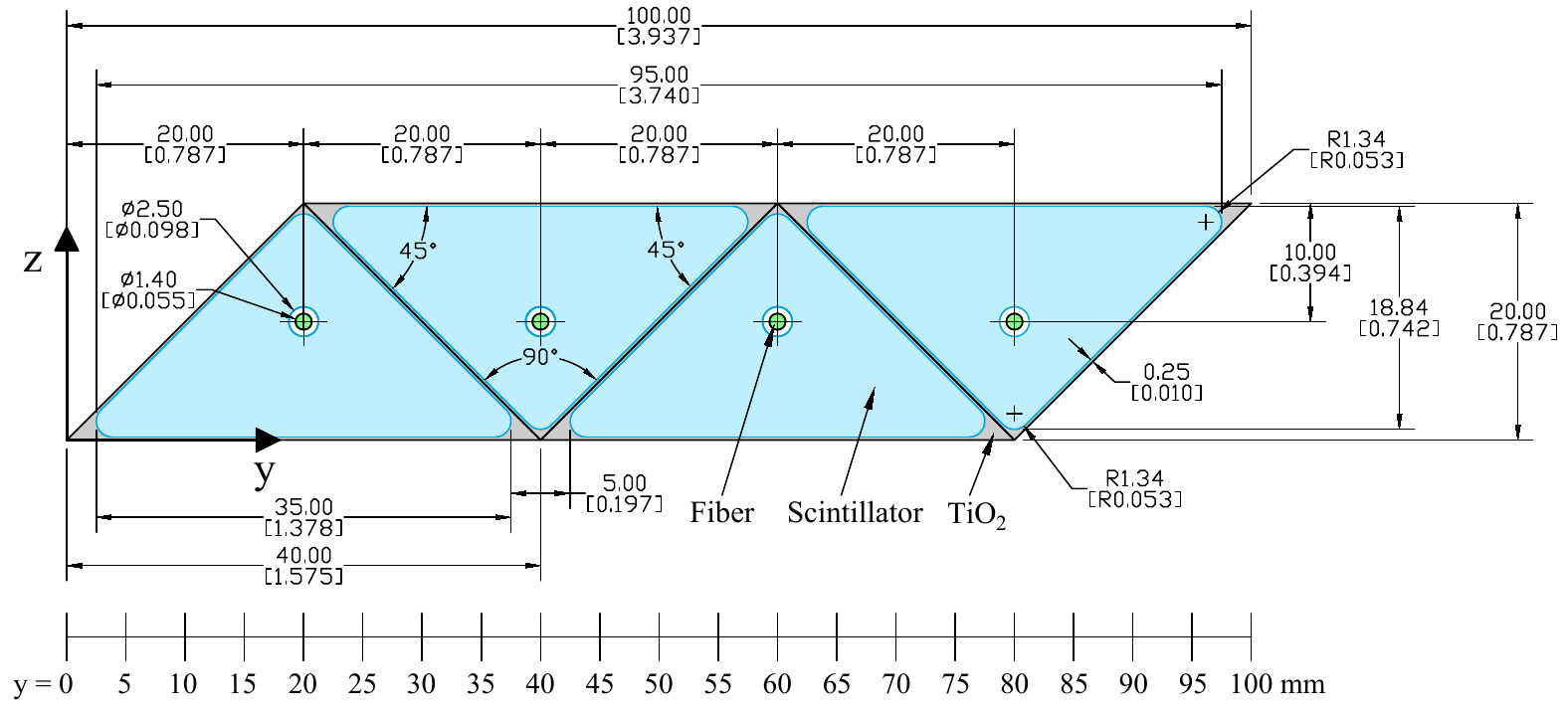}
\captionsetup{width=0.95\textwidth}
  \caption
  {Profile dimensions of a grouping of four triangular counters (quadcounter). 
  Although the exterior dimensions are 20.0\,mm height by 100.0\,mm width, the
  actual active scintillator dimensions are less due to `dead' PS/TiO$_2$ coating material
  filling in the corners of the triangular counters.
  For example, the 5.0\,mm gap between the scintillator of adjacent counters of the same 
  orientation is clearly visible. Upper (lower) dimensions are in  mm (inch).}
\label{fig:counter-type1}
\end{figure}

The fundamental element of the hodoscope is a 90$^{\circ}$-45$^{\circ}$-45$^{\circ}$
triangular counter (Fig.\,\ref{fig:counter-type1}) with a nominal base of 40\,mm length and 
a design height of 20\,mm. A channel of 2.5\,mm nominal diameter, into which
WLS fibers are inserted, is centered 10\,mm above the base.  The counters tested 
were extruded at the Fermilab NICADD facility \cite{nicadd} and consist of a 
polystyrene (PS) base doped with 1\% PPO, 0.03\% POPOP, and coated with a thin 
(nominally 0.25\,mm thick) PS/TiO$_2$ reflective layer.  The scintillator at the 
counter corners was rounded with the nominal radii shown in the figure.  

\begin{figure}
\centering
\includegraphics[width=0.50\columnwidth]{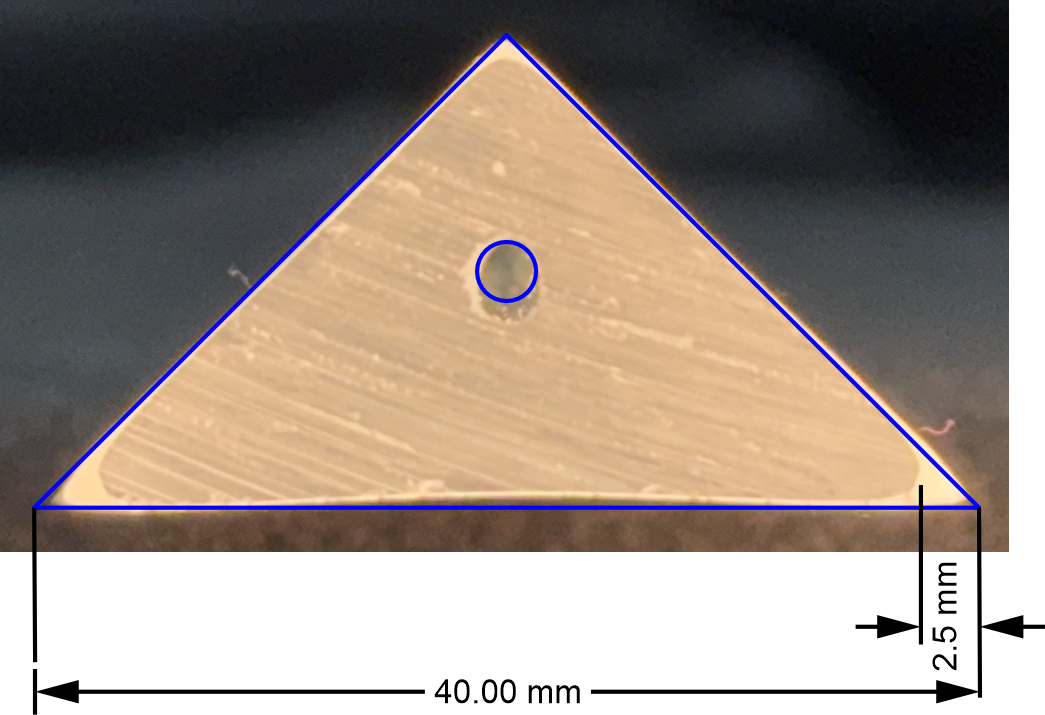}
\captionsetup{width=0.50\textwidth}
\caption{Photograph of the end of an extrusion after being rough cut with the desired 
counter profile superimposed. The extra fill of the co-extruded TiO$_2$
at the corners is clearly visible.}
\label{fig:extrusion-photo}
\end{figure}

The counters were grouped into arrays of four called quadcounters for 
readout purposes. The corners of every other counter touched without gaps. 
However, as shown in Fig.\,\ref{fig:counter-type1} and in the photo of an 
actual extrusion in Fig.\,\ref{fig:extrusion-photo}, the PS/TiO$_2$ reflective 
coating partially filled in the extrusion corners, resulting in a 
5.0\,mm gap of dead (non-scintillating) material in which a charged  particle 
produced light in only the counter above the gap, preventing light-sharing 
interpolation from being done and adversely impacting the resolution in that region. 

\section{Determining the Impact Position from Light Sharing}
\begin{figure}
\centering
\includegraphics[width=0.65\textwidth]{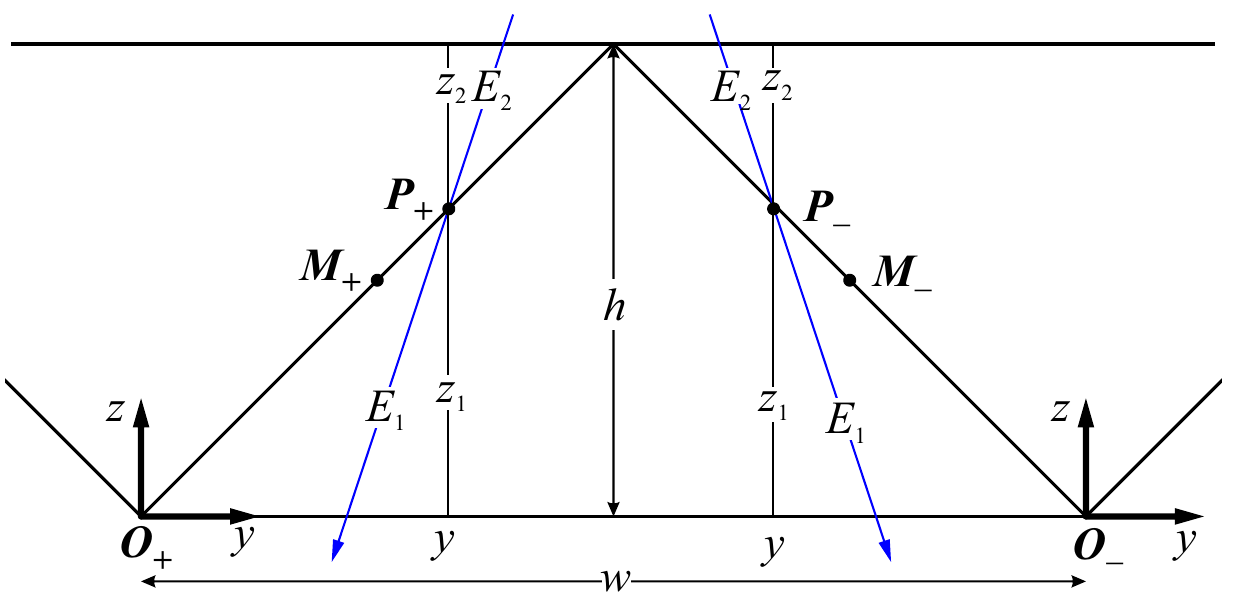}
\captionsetup{width=0.65\textwidth}
\caption{The counter coordinate system showing two charged tracks (in blue) traversing 
the inverted counters to the left and right of a central upright counter.
The $y$ and $z$ coordinates given in Eqs.\,\ref{eq:y} and \ref{eq:z} refer to the 
points $\boldsymbol{P_+}$ and $\boldsymbol{P_-}$ in which the charged particles pass 
the boundary between two counters and are with respect to the halfway points between 
the counters: $\boldsymbol{M_+}$ and $\boldsymbol{M_-}$ in this figure.  
}
\label{fig:geometry}
\end{figure}

We wish to find the coordinates of a charged particle going through a plane 
of triangular counters, determining them for an isosceles triangular geometry 
of arbitrary width $w$ and height $h$.  We assume that the angle of incidence 
of the particle is such that at most two counters are traversed. The geometry 
is shown in Fig.\,\ref{fig:geometry}.
The interpolated coordinates of the point at which a charged particle passes 
the boundary between two adjacent counters, $\boldsymbol{P_{\pm}}$,
relative to the origins set midway
between the two counters, $\boldsymbol{M}_{\pm}$ in Fig.\,\ref{fig:geometry}, 
has the following values:
\begin{align}
y_{\pm} & = \pm\frac{w}{4}\frac{E_1 - E_2}{E_1 + E_2}, \label{eq:y}\\
z_{\pm} & =     \frac{h}{2}\frac{E_1 - E_2}{E_1 + E_2}, \label{eq:z}
\end{align}
where $E_1$, $E_2$, are the lengths traversed by the charged particle,  
as shown in Fig.\,\ref{fig:geometry}, $w$ is the counter base length, $h$ is the
counter height, 
and the $\pm$ subscripts for the $y$ and $z$ coordinates in Eq.\,\ref{eq:y} 
and \ref{eq:z} refer to the coordinate origins $M_{\pm}$ in 
Fig.\,\ref{fig:geometry}.\footnote{Note that Eq.\,\ref{eq:y} and
Eq.\,\ref{eq:z} are true for isosceles triangles of any kind.} Note that, 
although all of the tracks used in this analysis were incident normal to the 
face of the quadcounter, Eqs.\,\ref{eq:y} and \ref{eq:z} are true irrespective 
of the incident angle of the tracks, as long as they do not traverse more than two counters.

To estimate the charged particle path lengths used in Eq.\,\ref{eq:y} and 
Eq.\,\ref{eq:z}, $E_1$ and $E_2$, can be replaced by the associated photoelectron 
yield (PE) proxies, $N_1$ and $N_2$, respectively. Assuming Poisson statistics, the 
uncertainties in the interpolated $y$ and $z$ positions of those two equations are:
  \begin{align}
\sigma_y & = \frac{\Delta z}{S} = \frac{w}{2}\sqrt{\frac{N_1N_2}{(N_1 + N_2)^3}}, \label{eq:dx} \\
\sigma_z & = h\sqrt{\frac{N_1N_2}{(N_1 + N_2)^3}} \label{eq:dy},
\end{align}
where $S = \Delta z / \Delta y$ in Eq.\,\ref{eq:dx} is the slope of the junction 
between the triangular counters (equal to $S = h/(w/2)$ for the geometry in 
Fig.\,\ref{fig:geometry}). 
\begin{figure}
\centering
{\includegraphics[width=0.65\textwidth, valign=c]{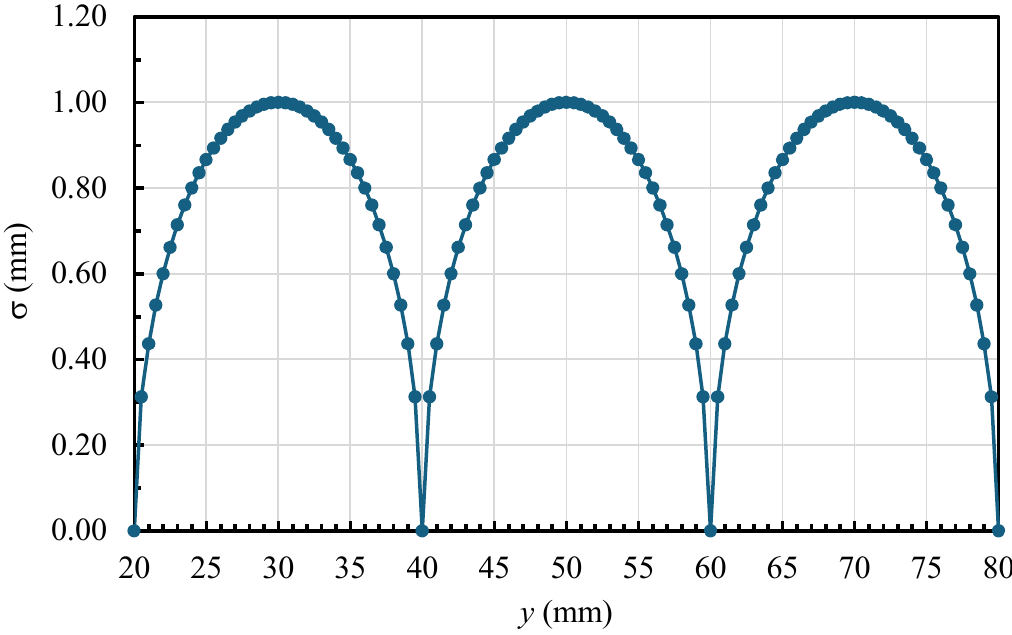}}
\captionsetup{width=0.65\textwidth}
\caption{The theoretical $y$-position uncertainty for charged particles 
traversing the three junction regions between the four counters in a quadcounter
with perfectly shaped triangular counters (no dead regions), 
assuming 5\,PE/mm and Poisson uncertainties in the counter light 
yields. The average uncertainty is $\sigma_y = 0.76$\,mm.}
\label{fig:uncertainty_theoretical}
\end{figure}

Using Eq.\,\ref{eq:dx}, the uncertainties in 
$y$ for particles traversing a quadcounter with perfectly shaped isosceles 
triangles (no rounded corners) of height 20\,mm and base width 40\,mm are shown 
in Fig.\,\ref{fig:uncertainty_theoretical}, where we have assumed the charged 
particles impact the counters at normal incidence  and a photoelectron yield 
of 5 per mm (effectively, the PE yield achieved in the 
test beam as shown below). As expected, the best resolution occurs at
the apexes of the triangular counters, whereas the worst occurs halfway in-between.
The average uncertainty is $\sigma_y = 0.76$\,mm. 

\begin{figure}
\centering
\includegraphics[width=0.65\textwidth]{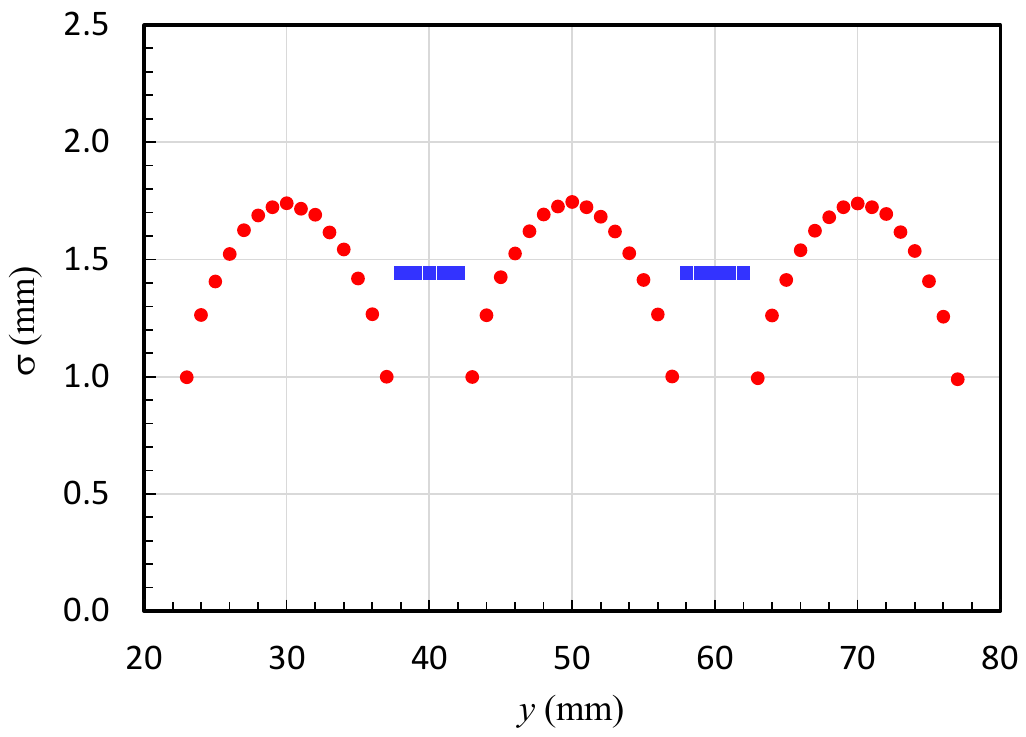}
\captionsetup{width=0.65\textwidth}
\caption{The $y$ position resolution as a function of $y$ from the Monte
Carlo simulation of the quadcounter shown in Fig.\,\ref{fig:counter-type1}.  
Units are mm. The red points represent data in which two 
adjacent counters had non-zero PE values, allowing for interpolation between
the two counters. (There are only three shared surfaces between the four
triangular counters.)  The blue points represent data where the track passed through the 
dead region between counters and hence interpolation was impossible, hence 
$\sigma = 5.0\,{\rm mm}/\sqrt{12} = 1.44$\,mm has been used for the dead region.}
\label{fig:simulation}
\end{figure}

\begin{figure}
\centering
\begin{minipage}[t]{3.2in}
\includegraphics[width=0.95\textwidth]{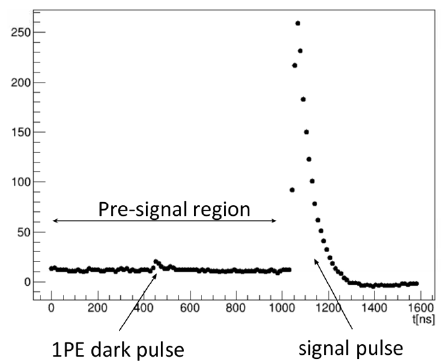}
\captionsetup{width=0.95\textwidth}
\caption{Typical time profile of an event.  The readout samples each SiPM output
every 12.5\,ns, and places the signal pulse near the end of the record so the
pre-signal region is observed, allowing dark pulses to be recorded for calibration
purposes.}
\label{fig:event}
\end{minipage}
\begin{minipage}[t]{3.2in}
\raisebox{0.15in}{\includegraphics[width=0.95\textwidth]{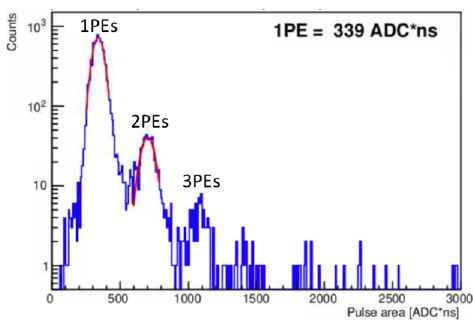}}
\captionsetup{width=0.95\textwidth}
\caption{Typical photoelectron (PE) spectrum from dark pulses in the pre-signal region.}
\label{fig:pe_spectrum}
\end{minipage}
\end{figure}

\section{Simulation Results of the Position Resolution}
A Monte Carlo simulation estimated the expected resolution of
a hodoscope with the design shown in Fig.\,\ref{fig:counter-type1}.
Charged particles were incident on the hodoscope at 1\,mm steps with a mean light
yield of 5 photoelectrons per mm when passing through the PS scintillator. The 
light yield in each counter was determined using a Gauss + Landau distribution 
whose parameters were extracted from the test-beam data, scaled by the amount of 
scintillator traversed by the charged particle.\footnote{We have found that
a Gauss + Landau distribution fits our data better than a pure Landau fit. 
See Ref.\,\cite{testbeam}.} 

The results are shown in Fig.\,\ref{fig:simulation}, where the red points
correspond to regions where the charged particle went through two adjacent
counters, allowing light sharing interpolation to be done. Again, as expected the best
resolution occurs when the charged particle passes near the dead region at the apex of one of
the triangular counters. At the 5.0\,mm of dead PS/TiO$_2$
at the corners of the triangles, where only one counter is hit, a resolution
(the blue points in Fig.\,\ref{fig:simulation}) of $\sigma = 5.0$\,mm/$\sqrt{12} = 1.44$\,mm
has been used.  The average resolution 
elsewhere is $\sigma_y = 1.48$\,mm.  The reason that the resolution is close to
a factor of two worse than the theoretical resolution shown in Fig.\,\ref{fig:uncertainty_theoretical}
is due to the replacement of the Poisson light yield distribution
with the measured Gauss + Landau distribution, which is wider and has a long tail,
and the omission of the gap region resolution points.

\section{Test Beam Results}
Four quadcounters were fabricated at the University of Virginia and taken to 
the Fermilab Meson Test 
Beam Facility \cite{ftbf} to measure their position resolution, among other 
aspects of their performance such as the photoelectron yield. 

\subsection{Quadcounter Description}

The triangular extrusions were fabricated at the Fermilab NICADD Extrusion Line 
Facility \cite{nicadd}.  The polystyrene base of each counter was STYRON 665 W. 
The primary dopant was 2,5-diphenyloxazole (PPO, 1\% by weight). The secondary 
dopant was 1,4-bis (5-phenyloxazol-2-yl) benzene (POPOP, 0.03\% by weight). A co-extruded reflective 
coating of nominal 0.25\,mm thickness surrounded the core and was composed of a 
co-extruded mixture of polystyrene mixed with 30\% TiO$_2$. Each counter also 
had a channel of nominal 2.5 mm diameter into which wavelength-shifting (WLS) 
fibers were placed.  A photo of an extrusion end is shown in Fig.\,\ref{fig:extrusion-photo} 
superimposed on the nominal profile. The fiber hole is elliptically 
shaped and situated a bit lower than design and the PS/TiO$_2$ coating is uneven and thicker at the
corners.\footnote{Work is underway at Fermilab to procure better dies to improve the profile.} 
The extra dead PS/TiO$_2$ material at the corners adversely affects the
position resolution at the junction between two counters, as shown below.

\begin{figure}
\centering
\includegraphics[width=0.70\columnwidth]{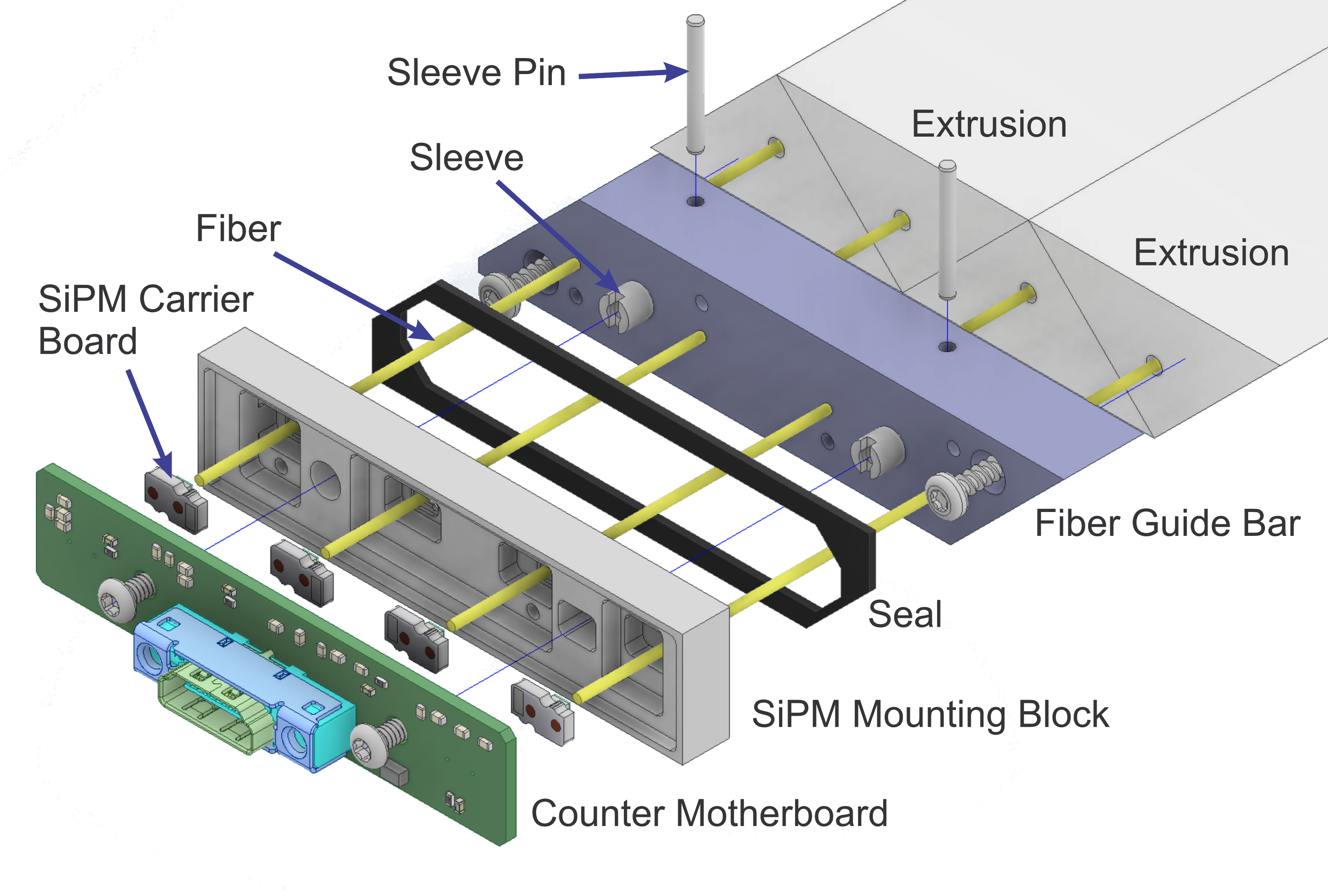}
\captionsetup{width=0.70\textwidth}
\caption{The quadcounter readout manifold. Each end of each quadcounter had identical readout manfiolds.}
\label{fig:manifold}
\end{figure}

Counters were assembled into full quadcounters at the University of Virginia. The counters 
were first glued into groups of four using 3M DP420 epoxy \cite{dp420}.  
The WLS fibers were then placed into the four quadcounter channels. The fibers were 1.4\,mm 
diameter Kuraray double-clad, non-S-type, Y11 doped with 175\,ppm K27 dopant 
\cite{kuraray}\cite{wls}. With the exception of one quadcounter, the fibers were not 
glued nor potted with any sort of epoxy or 
filler in the counter extrusion  channels, although such fillers have been shown to increase 
the light yield by about 50\% \cite{thesis-boi}.  At each end of a quadcounter an acetal 
fiber guide bar (see Fig.\,\ref{fig:manifold}) was glued to the extrusions using 3M DP100 
epoxy \cite{dp100}. At the same time the fibers were glued into funnel-shaped channels in the 
fiber guide bars using the same epoxy. The fibers, protruding from both ends of the 
quadcounters, were then cut off using a hot knife, after which the fiber guide bars were 
fly cut, which served to polish the fiber ends.

The quadcounter readout was based on that designed for the Mu2e experiment 
Cosmic Ray Veto detector \cite{thesis-boi}. The light from each fiber was detected using
$2.0 \times 2.0$\,mm$^2$ silicon photomultipliers (SiPMs): Hamamatsu model S13360-2050VE, with
a pixel size of $50\,\mu$m \cite{hamamatsu}. Each SiPM was mounted on a small,
$8.61 \times 5.61$\,mm$^2$, PC board called a SiPM Carrier Board. The SiPM Carrier Boards were
freely placed into small rectangular wells in an anodized aluminum piece called a SiPM
Mounting Block. Precision sleeves attached to the Fiber Guide Bars aligned the
SiPM Mounting Block to it, assuring that the fibers were precisely aligned to 
the SiPMs.  Finally, a small opaque PC board --- the Counter Motherboard --- interfaced
the SiPMs to the front-end readout boards.  Small pogopins on the Counter Motherboard
served to establish the electrical connection to the SIPMs and also gently pushed
them against the fibers. The Counter Motherboard, besides sending the signals 
from four SiPMs to the front-end readout boards, provided the SiPM bias.

The main component of the readout was the front-end board (FEB) \cite{feb}, 
each of which can serve up to 64 SiPMs.  
It provided bias to the SiPMs and amplified, shaped, digitized, zero-suppressed, 
and stored the signals coming from the SiPMs.
The four SiPMs of a Counter Motherboard connected to it via one HDMI cable.
Two FEBs were used, one at each end of the counters.
Each FEB uses commercial analog front-end (AFE \cite{afe}) chips to amplify and digitize 
the SiPM signals at 80~MS/s (mega samples per second).  Data continuously 
entering the AFEs were zero-suppressed by FPGAs and stored in a buffer memory, 
awaiting triggers to initiate the readout. The FEBs were powered using CAT6 
Ethernet cables connected to a readout controller (ROC). The ROC, which could 
also have collected data from the FEBs and served as the interface to the 
data acquisition (DAQ) computers, was not used in this fashion for this study.
Rather, the ROCs only supplied power, timing, and triggers, 
while data were sent directly from the FEBs to a DAQ computer. During data 
acquisition, triggers prompted the storage of pre- and post-trigger data 
for calibration and pedestal determination, with each event including 
127~ADC samples per channel for a total sampling duration of 1588\,ns.  

The photoelectron yield was determined on the off-line analysis as follows.
The signal from an event in one SiPM is shown in Fig.\,\ref{fig:event}.
In the pre-signal region, the data from each SiPM underwent a scan for dark pulses.
The pulses were pedestal subtracted and fitted using the following modified Gumbel
function \cite{gumbel}:
\[
  {\rm ADC}(t) = A \cdot e^{-\frac{t - \mu}{\beta} \cdot e^{-\frac{t-\mu}{\beta}}},
\]
where $\mu$ is the time of the pulse peak, the pulse height is $A/e$, and the pulse area 
is $A \cdot \beta$. The fitting process determined the dark count pulse areas, a plot of 
which is shown in Fig.\,\ref{fig:pe_spectrum}, where distinct photoelectron peaks can be
seen. This allowed the signal pulse areas to be converted into the number of photoelectrons.

\subsection{Experimental Setup}

\begin{figure}
\centering
\includegraphics[width=0.75\textwidth]{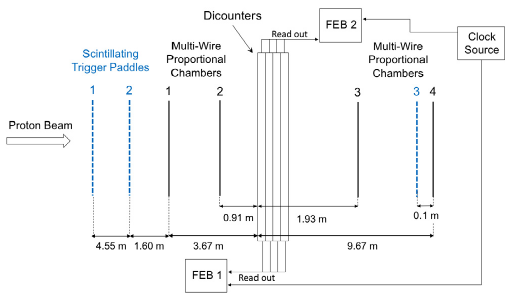}
\caption{Schematic of the Fermilab Meson Test Beam Facility setup, not drawn to scale.}
\label{fig:MTBF}
\end{figure}

\begin{figure}
\centering
\includegraphics[width=0.95\textwidth]{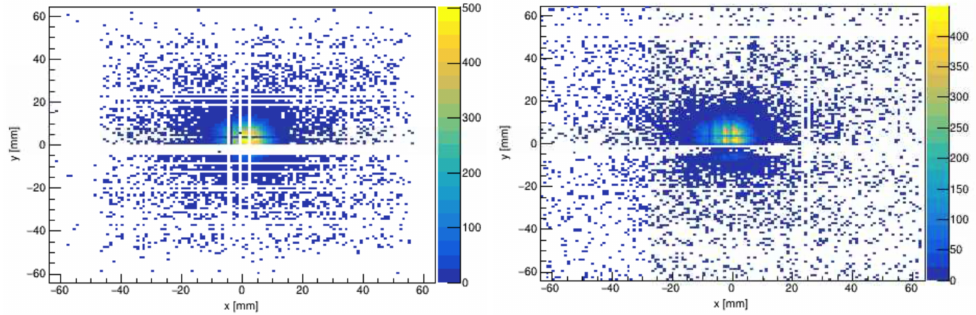}
\caption{MWPC hit maps of chamber 1 (left) and chamber 2 (right).
The $x$ ($y$) coordinate is horizontal (vertical). 
Regions of dead wires are clearly visible.}
\label{fig:MWPC_hitmap}
\end{figure}

The experimental setup is shown in Fig.\,\ref{fig:MTBF}. A 120\,GeV proton beam, 
delivered in roughly 4-s-long spills every minute, was incident on the apparatus.  
Multiwire proportional chambers (MWPCs) served to determine the position of the beam: two in 
front and two behind the quadcounter array. Unfortunately, the two rear MWPCs did 
not work, and there were regions in the upstream MWPCs that either did not work 
or were highly inefficient (see Fig.\,\ref{fig:MWPC_hitmap}), and there was 
insufficient time nor adequate  resources to find and fix the problems. Hence, 
the track reconstruction, which was only 30\% efficient, was compromised, nor 
was there any redundancy in the tracking, preventing the track-finding
resolution to be determined. 

The data acquisition consisted of a two-phase ``cycle'': a 220 second 
live-time period for data collection, followed by a 20 second period 
for data transfer to the DAQ PC.  Typical event rates were 3~kHz.  
Triggers were initiated by the coincidence of three scintillator counters 
(each $101.6 \times 101.6$\,mm$^2$ in size); two in front of the quadcounters 
and one behind.  The data from the MWPCs were taken using a different 
readout stream, triggered in exactly the same way as the quadcounter data, 
and merged together with the quadcounter data in the offline analysis.

\begin{figure}
\centering
\includegraphics[width=0.55\columnwidth]{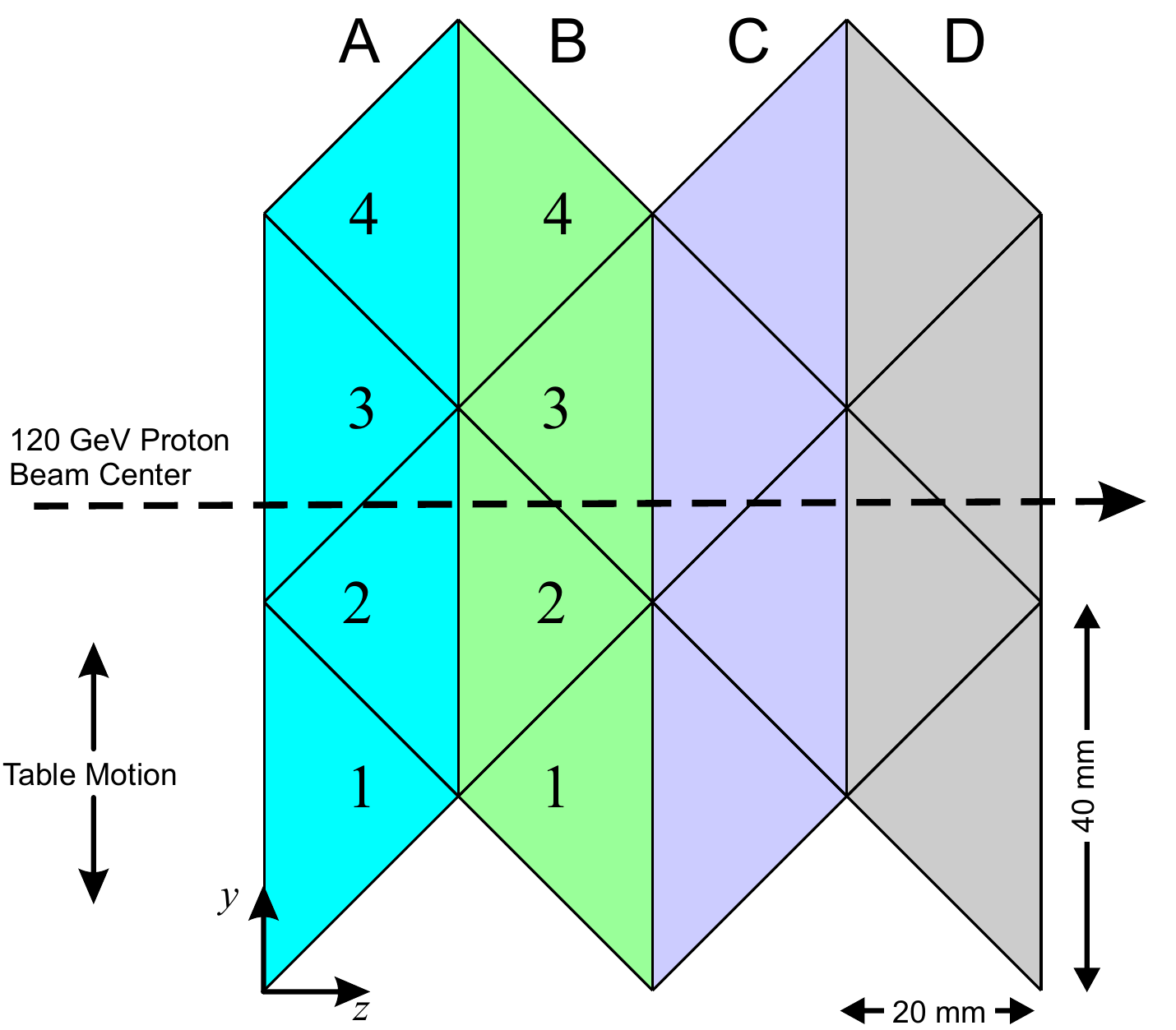}
\captionsetup{width=0.55\textwidth}
\caption{End view of the quadcounter setup showing the numbering scheme of
the two upstream quadcounters, which are the only ones whose 
results are presented in this paper.  The beam position was
fixed; the table on which the counters were mounted, moved
vertically ($\pm y$).}
\label{fig:counter_setup}
\end{figure}

Four quadcounters were placed on a fixture mounted on a table in the orientation 
shown in Fig.\,\ref{fig:counter_setup}. The first three quadcounters 
(A, B, and C in Fig.\,\ref{fig:counter_setup}) were 3.35-m long; the fourth 
was 1.0-m long. One of the long counters (C in Fig.\,\ref{fig:counter_setup}) 
had fibers potted with Solaris silicone \cite{solaris}, but unfortunately was 
damaged in transit.  The table could be moved in the vertical 
($\pm y$) direction,  allowing the photoelectron yield to be determined as a 
function of the impact position on the triangular counters.  The fixture 
could also be rotated. We report here only on the data taken with the proton 
beam at normal incidence and only on data from the two upstream quadcounters.

\subsection{Data Analysis}

Data were taken with the beam center positioned from $y = 15-84$\, mm (see
Fig.\,\ref{fig:counter-type1} defining the coordinate system), with 
the table moved in 3\,mm increments, as the beam FWHM was approximately 3\,mm.  
The beam was centered on $x = 0.8$\,m from the near-side quadcounter end. 
Only analysis from data taken from that end are reported here.

\begin{figure}
\centering
\includegraphics[width=0.45\textwidth]{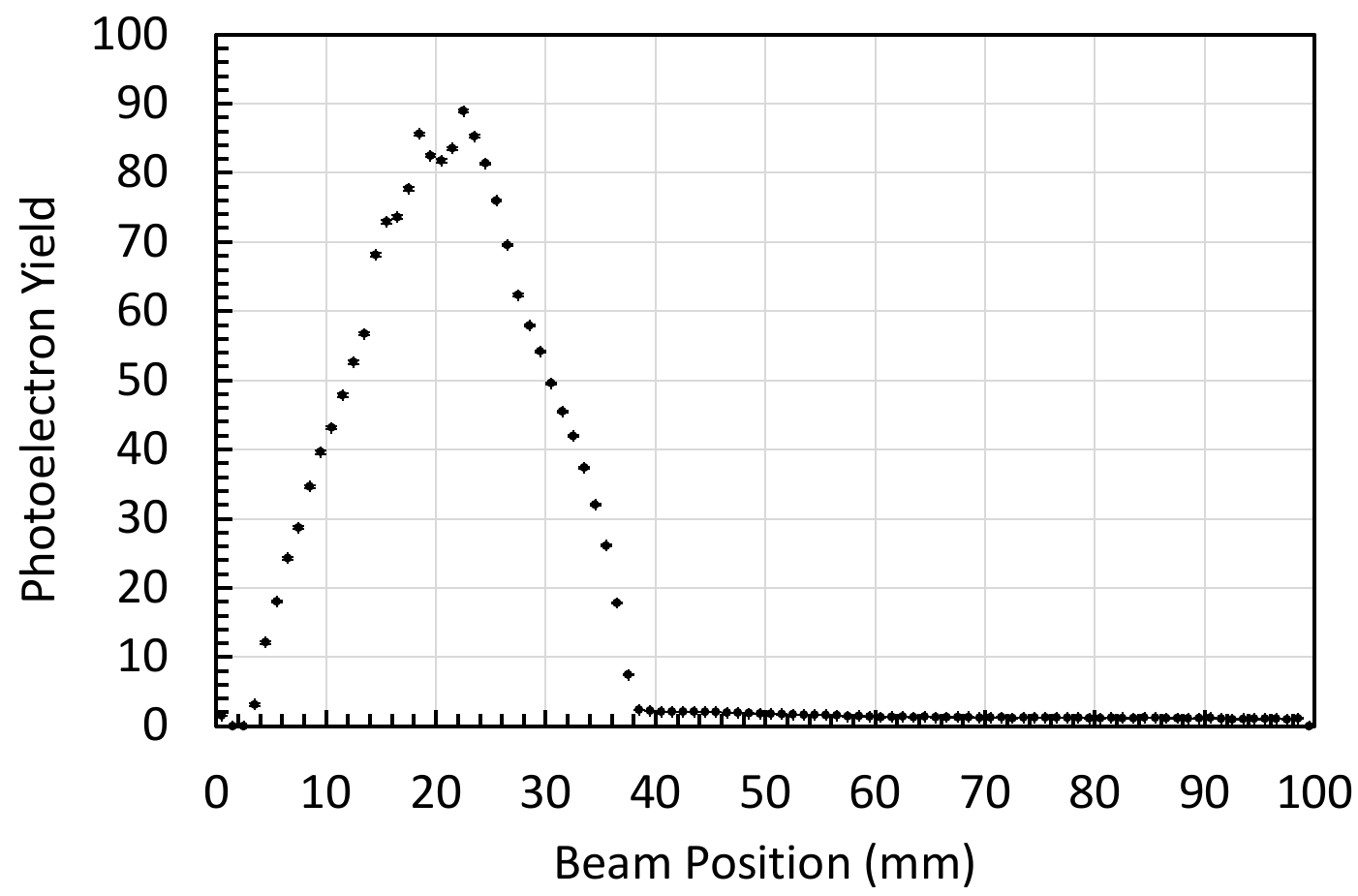}
\includegraphics[width=0.45\textwidth]{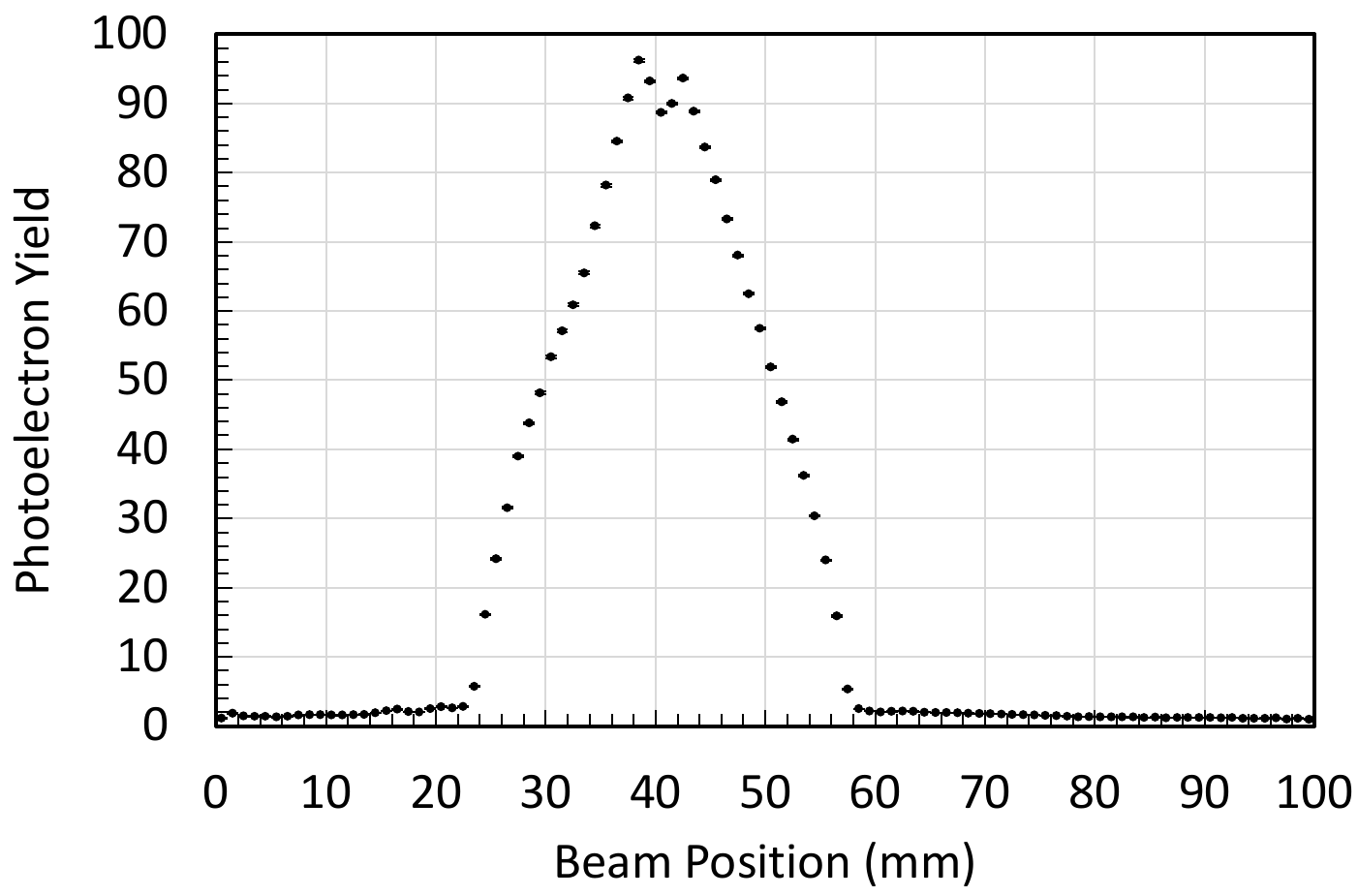}
\includegraphics[width=0.45\textwidth]{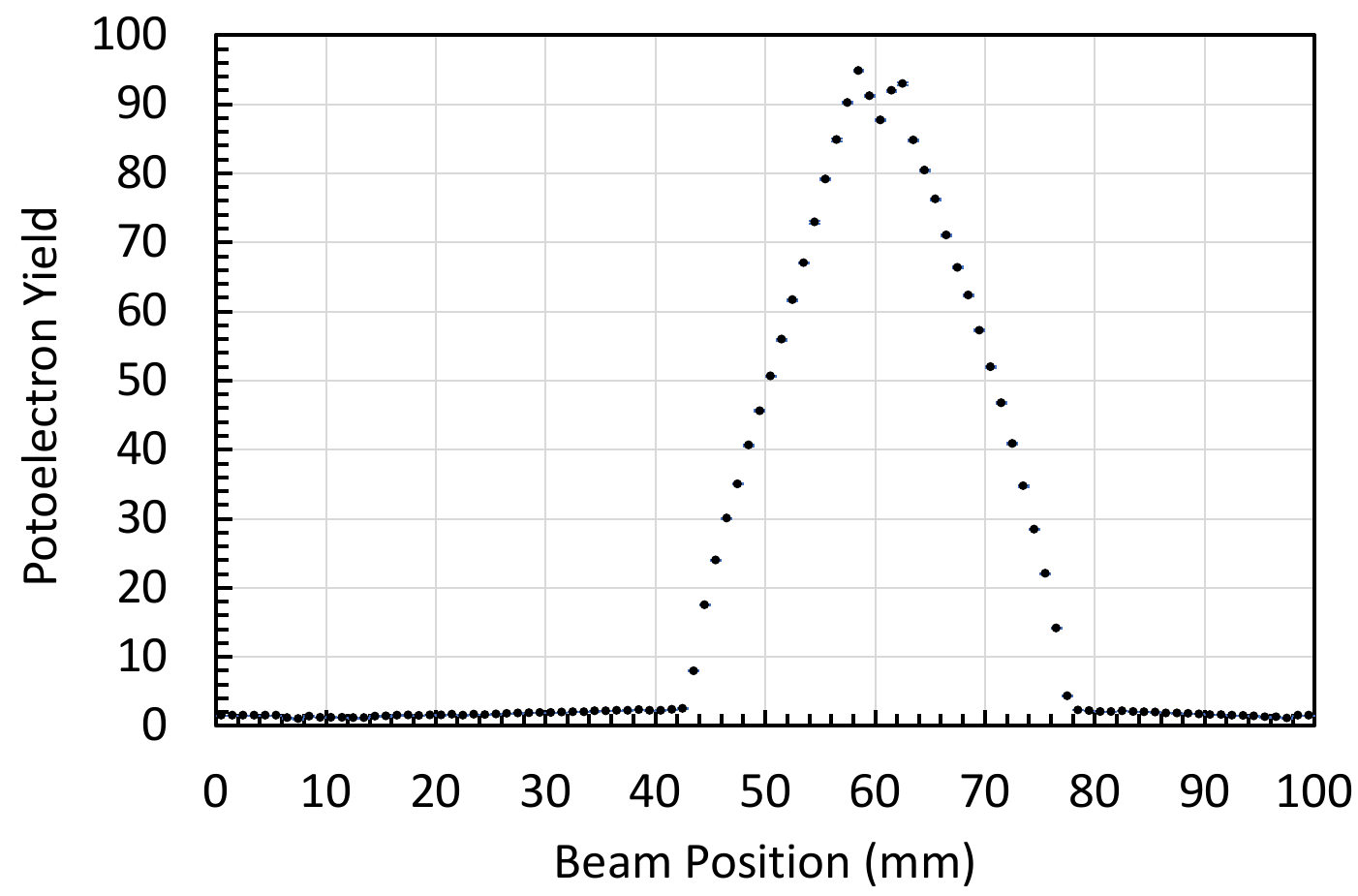}
\includegraphics[width=0.45\textwidth]{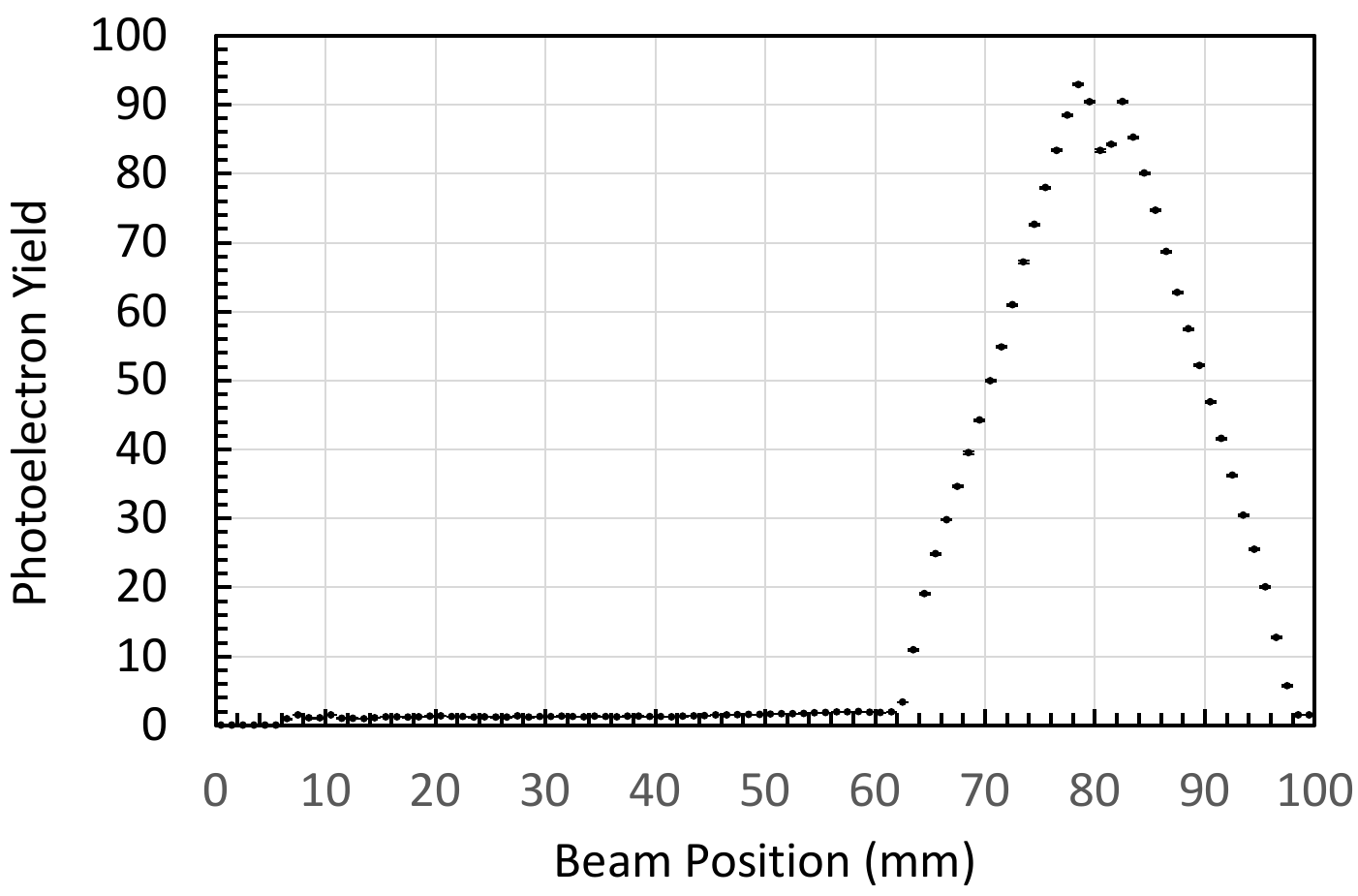}
\caption{The average photoelectron yields of the front quadcounter counters
as a function of the vertical $y$ positions as measured by the MWPCs.
The counters are numbered from 1 to 4, top left to bottom right 
(see Fig.\,\ref{fig:counter_setup}).  
Dips at the peaks are due to the fiber channel.  Units are mm.}
\label{fig:counter_profiles}
\end{figure}

Figure\,\ref{fig:counter_profiles} shows the average photoelectron yields as a function of
the $y$ position of the four triangular counters in the front quadcounter. There is a slight
non-zero PE yield when the beam was away from the counters that was due to SiPM noise 
and multiple particles in the beam.  (No threshold on the PE yield was used for these plots.)
The yield then increases linearly as the beam traverses more scintillator, peaking 
slightly below 100 photoelectrons. Dips can be seen at the peaks, which are due to
less scintillator at the fiber channels. The average photoelectron peak separation is 20.3\,mm 
and the average width at the base is 35.3\,mm, which is consistent with the average 
scintillator width of the triangular counters shown in Fig.\,\ref{fig:counter-type1} and the
2.5\,mm of dead material at the counter corners (see Fig.\,\ref{fig:extrusion-photo}). 
The average FWHM of the counter photoelectron yields is 20.6\,mm.

Data were taken with the beam center positioned from $y = 15-84$\, mm (see
Fig.\,\ref{fig:counter-type1} defining the coordinate system), with 
the table moved in 3\,mm increments, as the beam FWHM was approximately 3\,mm.  
The beam was centered at $x = 0.8$\,m from the near-side quadcounter end. 
Only analysis from data taken from that end are reported here.

Tracks were fitted using the two upstream MWPCs as the two downstream MWPCs
did not work. In each MPWC view ($x$ and $y$) events with more than two
hits wires separated by at least one wire were not used. A timing cut of
283\,ns ($30{\times}$ the 106\,MHz RF time) was applied to assure that the MWPC hits corresponded to the same
proton. However, because of the dead regions in the MWPCs 
(see Fig.\,\ref{fig:MWPC_hitmap}), we could not be sure that there were 
no extra tracks in any of the events.

Figure\,\ref{fig:counter_profiles} shows the average photoelectron yields 
as a function of the $y$ position of the four triangular counters in the 
front quadcounter. There is a slight non-zero PE yield when the beam was 
away from the counters that was due to SiPM noise and multiple particles 
in the beam.  (No threshold on the PE yield was used for these plots.)
The yield then increases linearly as the beam traverses more scintillator, 
peaking  slightly below 100 photoelectrons. Dips can be seen at the peaks, 
which are due to less scintillator at the fiber channels. The average 
photoelectron peak separation is 20.3\,mm and the average width at the 
base is 35.3\,mm, which is consistent with the average scintillator 
width of the triangular counters shown in Fig.\,\ref{fig:counter-type1} 
and the 2.5\,mm of dead material at the counter corners (see 
Fig.\,\ref{fig:extrusion-photo}). The average FWHM of the counter 
photoelectron yields is 20.6\,mm.

\begin{figure}
\centering
\includegraphics[width=0.49\textwidth]{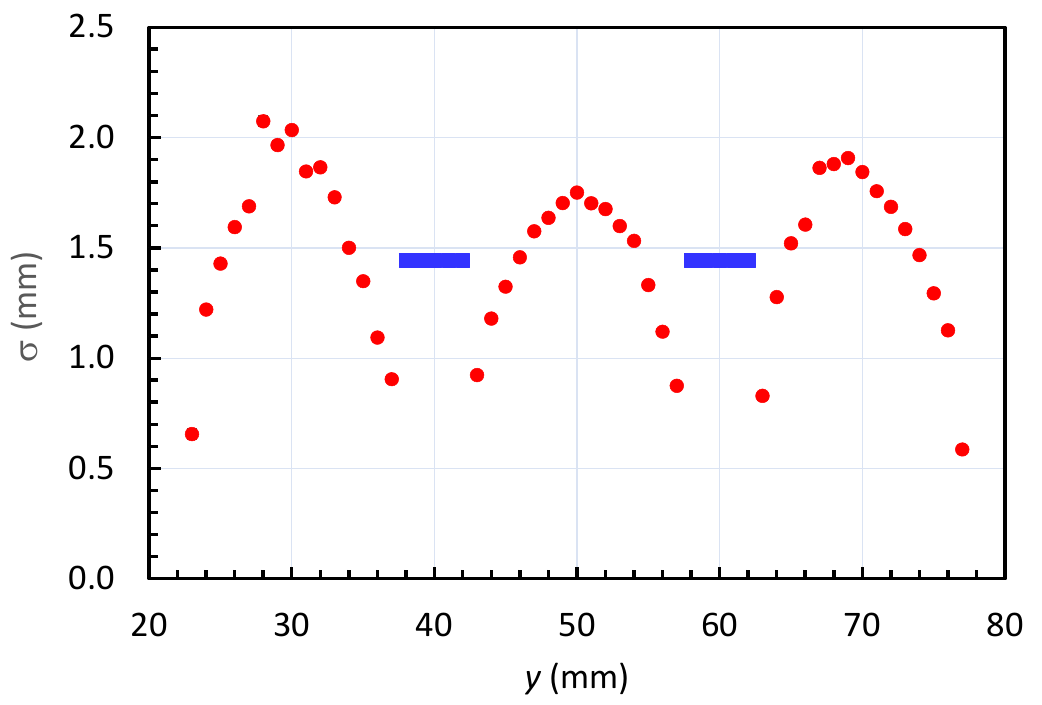}
\includegraphics[width=0.49\textwidth]{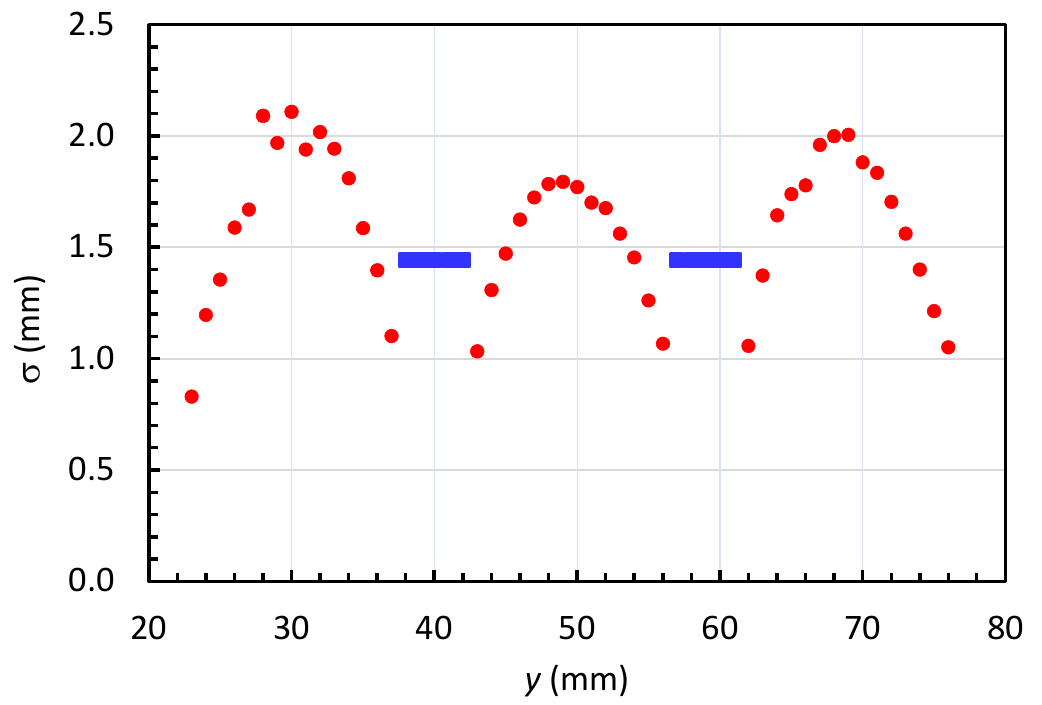}
\caption{The $y$ position resolution ($\sigma_y$) for
quadcounter A (left) and B (right), as a function of $y$ as measured by the MWPCs.  
Units are mm. The red points represent data in which two adjacent counters
had PE values of three or more, allowing interpolation to be done. The blue 
points represent the region where the track passed through the gap between 
counters and only the counter in between had non-zero PE values, preventing
the interpolated position from being determined. A resolution of $\sigma_y = 1.44$\,mm
is used in that region, as determined by assuming uniform illumination of the 5.0\,mm
gaps.}
\label{fig:y-resolution}
\end{figure}

The $y$-position resolution, $\sigma_y$, vs $y$, is shown in Fig.\,\ref{fig:y-resolution}
for both quadcounter~A and quadcounter~B.
The data were required to have a valid track, a minimum of 3\,PEs in each
counter, peak pulse times within 20\,ns of each other,
and the data for quadcounter A (B) required a combined light yield
between 50 and 150\,PEs in quadcounter B (A). This last requirement served 
to eliminate extra charged particle tracks that may have gone through
the dead, or inefficient, MWPC regions.

To determine the position resolution of the quadcounter (for normally incident
protons) two different cases were considered.
The first case required that the MWPC track did not point to the dead gap 
between counters, allowing interpolation to be done using Eq.\,\ref{eq:y}.  
The difference between the projected track position and 
the quadcounter interpolated position was 
plotted and fitted to a Gaussian distribution to determine $\sigma_y$.
These are the red circles in Fig.\,\ref{fig:y-resolution}.
Note that the unknown MWPC track resolution is included in the uncertainties.

In the second case, for all tracks that: (1) projected to the dead gap between two counters 
of the same orientation (i.e., the gap between counters 1 and 3 and between 
counters 2 and 4 in Fig.\,\ref{fig:counter_setup}); and (2) that only the 
counter above the dead region gap had a non-zero photoelectron value 
(i.e., counters 2 and 3 in Fig.\,\ref{fig:counter_setup}), the resolution was
fixed to $\sigma_y = 5.0$\,mm$/\sqrt{12} = 1.44$\,mm.  That result is shown as the 
blue squares in Fig.\,\ref{fig:y-resolution}. 

As expected, the resolution is worse at the 
locations halfway between the apexes of the adjacent counters and best near the 
triangular-counter apexes. At the two apexes of the triangular counters
the resolution is worse since the track only traverses one counter and
interpolation is impossible for the 5\,mm of dead material between counters.
The average $y$-position resolution over the entire range of $y$ values in 
which interpolation is possible in Fig.\,\ref{fig:y-resolution}
is $\sigma_y = 1.48$\,mm (1.59\,mm) for quadcounter A (B). 
This is somewhat worse than what is expected
in the same range of $y$ values from the simulation results reported above.
Note that if one assumes an MWPC track resolution of 
$\sigma_y = 0.5$\,mm, which is what was measured in a previous test-beam experiment done
at the Fermilab Meson Test Beam facility (see Ref\,\cite{testbeam}), 
then we obtain an average resolution, where interpolation was possible,
of $\sigma_y = 1.39$\,mm. (1.51\,mm) for quadcounter A (B).

The resolutions determined at the three counter junctions --- 
between counters 1 \& 2, counters 2 \& 3, and counters 3 \& 4 --- 
differ somewhat for reasons that could not be determined, despite much effort.
Counters 1 and 4 for both quadcounter A and B had more backgrounds than
counters 2 and 3, which adversely affected their resolutions.  
The origin of these differences is not known, but could be due to, for example, 
different beam conditions when the data were taken. The data were taken
parasitically with another experiment whose detectors, placed upstream
of ours, were being moved in and out of the proton beam at times that
were not provided to us.

\section{Conclusions}

We have found that a hodoscope with triangular counters with a nominal fiber pitch of 20\,mm has
an average resolution between $\sigma = 1.5$\,mm and 1.6\,mm, obtained using photoelectron interpolation
between adjacent counters. Improving the light yield would improve the resolution. 
An easy $\sim$50\% increase in light yield would come from potting
the fibers.  A larger diameter fiber would also do so; see, for example Ref.\,\cite{testbeam}.
The penalty in using the former stratagem is it makes the quadcounter fabrication 
more difficult. The penalty of the latter stratagem is the increased cost.
On the other hand, increasing the light yield would allow smaller and thinner 
counters to be fabricated, if the material thickness is a concern. There are 
efforts at Fermilab and elsewhere to improve the light yield by improving the 
reflectivity of the counter reflective coating at shorter wavelengths. 

Other improvements include addressing the poorer position resolution at the 
counter corners due to the dead PS/TiO$_2$ material. Improved extrusion dies
at the Fermilab NICADD facility are being considered that should greatly reduce
the dead corner material.

Several detectors of similar design have been fabricated for various experimental efforts.
For example, the MINERvA neutrino experiment has reported on the performance of a similar
design in Ref.\,\cite{minerva}.\footnote{MINERvA reports a resolution of 3.1\,mm using
a triangular hodoscope with a 33\,mm base length and 17\,mm height.} 
A detector employing quadcounters of the design described 
here, and fabricated at the University of Virginia, is being used by the NAUM collaboration
to probe the structure of the interior of the Temple of Kukulk\'{a}n (El Castillo) at Chich\'{e}n Itz\'{a} 
using cosmic-ray muons. A proposed upgrade to the Mu2e experiment at Fermilab is also considering 
using similar counters \cite{mu2e2} as well as a proposed effort to scan the interior of
the Pyramid of Khufu in Giza, Egypt \cite{egp}.

\section{Acknowledgments}
We thank University of Virginia technicians Wayne Farrell and Eric Fernandez who were of 
invaluable help in fabricating the quadcounters. In addition, we are grateful for the 
vital contribution of the Fermilab staff of the Meson Test Beam Facility and the NICADD 
Extrusion Line Facility.  The work was supported in part by the U.S. Department of Energy 
under contract DE-SC0007838.

\phantomsection


\addcontentsline{toc}{section}{References}

\begin{thebibliography}{99}

\bibitem{nicadd}
A. Pla-Dalmau, A.D. Bross, V. Rykalin, ``Extruding Plastic Scintillator at Fermilab'',
2003 IEEE NSS Conference Record, FERMILAB-CONF-03-318-E.

\bibitem{ftbf}
E. Ramberg, Proceedings of the 2007 IEEE Nuclear Science Symposium and Medical
Imaging Conference (NSS/MIC 2007), 4436616, 2007.

\bibitem{dp420}
Scotch-Weld DP420, 3M, 3M Center, St. Paul, MN 55144, USA.

\bibitem{kuraray}
Kuraray America, Inc. 200 Park Ave. NY 10166 USA; 3-1-6, NIHONBASHI, CHUOKU, TOKYO 103-8254, JAPAN. 
http://kuraraypsf.jp/psf. (Accessed July 2017).

\bibitem{wls}
D. Coveyou, E.C. Dukes, R.C. Group, Y. Oksuzian, S. Roberts, and M. Solt,
``Performance of the Wavelength-shifting fiber upgrade for the Mu2e cosmic-ray veto detector'',
JINST {\bf 18} T05004 (2023).
and
E.C. Dukes, P.J. Farris, R.C. Group, T. Lam, Y. Oksuzian and D. Shooltz,
``Performance of wavelength-shifting fibers for the Mu2e cosmic ray veto detector'',
JINST {\bf  13} P12028 (2018).

\bibitem{thesis-boi}
S. Boi, ``Design and Fabrication of a Novel Large-area, High-efficiency Cosmic Ray Veto Detector 
for the Mu2e Experiment,''
PhD thesis, University of Virginia, 2021, 
\url{https://doi.org/10.18130/0nf5-sw49}.

\bibitem{dp100}
Scotch-Weld DP100, 3M, 3M Center, St. Paul, MN 55144, USA.

\bibitem{hamamatsu}
Hamamatsu Photonics K.K. 325-6 Sunayama-cho, Naka-ku, Hamamatsu City,
Shizuoka Pref. 430-8587, Japan. Hamamatsu Corp. 360 Foothill Rd. Bridgewater,NJ 08807.

\bibitem{feb}
S. Hansen, P. Rubinov, TWEPP 2015 –Topical Workshop on Electronics for Particle
Physics, Lisbon, 2015.

\bibitem{afe}
AFE5807, Texas Instruments Incoporated, 12500 TI Blvd., Dallas, TX 75243, USA.

\bibitem{gumbel}
G. Gavalian, Waveform Fitting Algorithm (unpublished), 2017. http://hallaweb.jlab.
org/experiment/DVCS/arswavefit.pdf. (Accessed July).

\bibitem{solaris}
Solaris, Smooth-On Co.,
5600 Lower Macungie Road
Macungie, PA 18062.

\bibitem{testbeam}
A. Artikov {\it et al.}, ``Photoelectron yields of scintillation counters with embedded
wavelength-shifting fibers read out with silicon photomultipliers'',
Nucl.\ Instrum.\ Meth.\ A 890 (2018) 84.

\bibitem{minerva}
L. Aliag {\it et al.}, ``Design, calibration, and performance of the MINERvA detector'',
Nucl.\ Instrum.\ Meth.\ A 743 (2014) 130.

\bibitem{mu2e2}
K. Byrum {\it et al.}, ``Mu2e-II: Muon to Electron Conversion with PIP-II'',
(Mu2e collaboration), 2022 Snowmass Summer Study, arXiv:2203.08192 (2022).

\bibitem{egp}
A. Bross {\it et al.}, ``Tomographic Muon Imaging of the Great Pyramid of Giza'',
J. Adv.\ Instrum.\ Sci.\ 280 (2022).



\end{thebibliography}
\end{document}